\documentclass{article}
\usepackage{arxiv}

\usepackage[utf8]{inputenc}
\usepackage[T1]{fontenc}
\usepackage[UKenglish]{babel}
\usepackage{amsmath,amssymb,amsthm}
\usepackage{booktabs,tabularx}
\usepackage{graphicx}
\usepackage{microtype}
\usepackage{xcolor}
\usepackage[hyphens]{url}
\definecolor{VUB_blauw}{rgb}{0.1529,0.2667,0.5529}
\usepackage{hyperref}
\hypersetup{
  colorlinks=true,
  citecolor=VUB_blauw,
  filecolor=VUB_blauw,
  linkcolor=VUB_blauw,
  urlcolor=VUB_blauw,
  pdftitle={Steiner Tree Parameterized by Multiway Cut: Single-Exponential Time and Polynomial Space},
  pdfauthor={Lukas Van de Voorde; Brecht Verbeken; Vincent Ginis}
}
\usepackage[nameinlink,noabbrev]{cleveref}
\newcommand{\paperdate}{30 October 2026}
\title{Steiner Tree Parameterized by Multiway Cut:\\Single-Exponential Time and Polynomial Space}
\runningtitle{Steiner Tree and Multiway Cuts}
\author{%
  Lukas Van de Voorde\\
  \normalfont Data Analytics Lab\\
  \normalfont Vrije Universiteit Brussel (VUB)\\
  \normalfont Brussels, Belgium\\
  \normalfont\href{mailto:lukas.van.de.voorde2@vub.be}{\texttt{lukas.van.de.voorde2@vub.be}}\\
  \normalfont\small ORCID: \href{https://orcid.org/0009-0006-8797-0676}{0009-0006-8797-0676}%
  \And
  Brecht Verbeken\\
  \normalfont Data Analytics Lab\\
  \normalfont Vrije Universiteit Brussel (VUB)\\
  \normalfont Brussels, Belgium\\
  \normalfont\href{mailto:brecht.verbeken@vub.be}{\texttt{brecht.verbeken@vub.be}}\\
  \normalfont\small ORCID: \href{https://orcid.org/0000-0002-7506-3298}{0000-0002-7506-3298}%
  \And
  Vincent Ginis\\
  \normalfont Data Analytics Lab\\
  \normalfont Vrije Universiteit Brussel (VUB)\\
  \normalfont Brussels, Belgium\\
  \normalfont\href{mailto:vincent.ginis@vub.be}{\texttt{vincent.ginis@vub.be}}\\
  \normalfont\small ORCID: \href{https://orcid.org/0000-0003-0063-9608}{0000-0003-0063-9608}
}

\newcounter{algorithm}
\newenvironment{algorithmblock}{%
  \par\medskip\noindent\begin{minipage}{\linewidth}\small\hrule\medskip
}{\medskip\hrule\end{minipage}\par\medskip}
\makeatletter
\newenvironment{spacetable}{%
  \begin{center}\begin{minipage}{\linewidth}\def\@captype{table}%
}{\end{minipage}\end{center}}
\makeatother

\theoremstyle{plain}
\newtheorem{theorem}{Theorem}[section]
\newtheorem{lemma}[theorem]{Lemma}
\newtheorem{proposition}[theorem]{Proposition}

\theoremstyle{definition}
\newtheorem{definition}[theorem]{Definition}

\theoremstyle{remark}

\numberwithin{equation}{section}

\DeclareMathOperator{\OPT}{OPT}

\DeclareMathOperator{\STVal}{STVal}
\DeclareMathOperator{\ST}{ST}
\DeclareMathOperator{\Space}{Space}
\newcommand{\poly}{\operatorname{poly}}

\newcommand{\colourfamily}{\mathcal{H}}

\theoremstyle{plain}
\newtheorem{literaturetheorem}{Theorem}

\crefname{literaturetheorem}{Theorem}{Theorems}

\begin{document}
\raggedbottom
\renewcommand{\today}{\paperdate}
\bibliographystyle{plainurl}
\maketitle
\thispagestyle{plain}

\begin{abstract}
In the Steiner Tree problem, the goal is to find a minimum-weight tree
containing a given set of terminals in an edge-weighted graph.
Given a multiway cut of size \(s\), we propose a deterministic algorithm
running in \(2^{O(s)}\poly(n,b)\) time and polynomial space
(\(n\) vertices, non-negative \(b\)-bit integer weights).
A multiway cut for the terminals is a set of vertices whose deletion
leaves at most one terminal in each connected component.
This result answers a question of Jansen and Swennenhuis (ESA 2024).
Building on their component decomposition, we show that it suffices
to connect the selected cut vertices together with \(O(s)\) additional
terminals and attach the remaining terminals by shortest paths.
We use colour coding to optimize over the choice of these additional terminals.
\end{abstract}


\section{Introduction}\label{sec:introduction}

In this work, we resolve an open question of Jansen and Swennenhuis \cite{jansen2024} by giving a polynomial-space algorithm for the Steiner Tree problem whose running time is single-exponential in the size of a multiway cut for the terminals.

In the Steiner Tree problem, the input is an undirected graph \(G=(V,E)\) with non-negative edge weights and a set of terminals \(K\subseteq V\). The goal is to find a minimum-weight connected subgraph containing all vertices of \(K\). An optimum can be chosen to be a tree, called a \emph{Steiner tree} for \(K\). We write \(n=|V|\), \(m=|E|\), and \(k=|K|\). This work considers non-negative integer weights represented with at most \(b\geq1\) bits.

We study the Steiner Tree problem using a \emph{multiway cut} \(S\) for \(K\): a vertex set \(S\subseteq V\) such that each component of \(G-S\) contains at most one terminal. Terminals are allowed to belong to \(S\). We present an algorithm parameterized by \(s=|S|\). The size of a multiway cut can be much smaller than \(|K|\): in a star whose leaves are terminals, the centre vertex alone is a multiway cut.

Karp \cite{karp1972} proved that the Steiner Tree problem is NP-hard.
The classical Dreyfus--Wagner algorithm \cite{dreyfus1971} runs in
\(3^k\poly(n,b)\) time, where \(k=|K|\). Fuchs et al.
\cite{fuchs2007} improve this to
\((2+\varepsilon)^k\poly_{\varepsilon}(n,b)\) time for every fixed
\(\varepsilon>0\). Both algorithms use exponential space.
Fomin et al. \cite{fomin2019} combine single-exponential time in
\(k\) with polynomial space (see Theorem~\ref{thm:solver}). We use their algorithm in our approach.

Jansen and Swennenhuis \cite{jansen2024} give an algorithm using \(2^{O(s\log s)}\poly(n,b)\) time and polynomial space, as well as an algorithm that runs in \(2^{O(s)}\poly(n,b)\) time but uses exponential space. For each chosen set of cut vertices used by the solution, their polynomial-space algorithm enumerates connecting systems that describe how subtrees connect these vertices. Counting these systems using Cayley's formula gives the \(2^{O(s\log s)}\) factor. They explicitly ask whether single-exponential time and polynomial space can be achieved simultaneously \cite[Section~5]{jansen2024}. This is the main result of this work.

\begin{theorem}\label{thm:main}
There exists a deterministic algorithm that, given an undirected graph
\(G=(V,E)\) with weight function \(w:E\to\mathbb{N}_0\), a set of terminals
\(K\subseteq V\), and a multiway cut \(S\) for \(K\), finds a minimum-weight
Steiner tree for \(K\) in \(G\), if it exists, in
\(2^{O(s)}\poly(n,b)\) time using polynomial space.
\end{theorem}

Since terminals may belong to a multiway cut, \(S=K\) is a valid choice.
Theorem~\ref{thm:main} therefore yields a
\(2^{O(k)}\poly(n,b)\)-time algorithm using polynomial space.

The algorithm of Chen, Liu, and Lu \cite{chen2009} requires the cut to
avoid terminals. We can handle this by forming an auxiliary graph
\(G^+\): for each terminal \(t\in K\), add a new leaf \(t'\) adjacent
to \(t\), and use \(K'=\{t':t\in K\}\) as the new terminal set.
A multiway cut for \(K'\) in \(G^+\) is also a multiway cut for \(K\)
and may contain original terminals. Since \(G^+\) has at most \(2n\)
vertices, the running-time and space bounds of Chen, Liu, and Lu
remain unchanged.

A minimum multiway cut can therefore be found in
\(4^\sigma\poly(n)\) time and polynomial space,\footnote{Polynomial space follows from bounded-depth
branching on important separators; see Jansen and Swennenhuis
\cite[p.~76:2, footnote~1]{jansen2024}.} where \(\sigma\) is the minimum
multiway cut size. Combining this with Theorem~\ref{thm:main} gives
\(2^{O(\sigma)}\poly(n,b)\) time and polynomial space if no multiway
cut is supplied.

\paragraph*{Our approach.}
We build on the component decomposition of Jansen and Swennenhuis
\cite{jansen2024}. For each chosen set of cut vertices used by the solution,
we reduce the optimization to choosing \(O(s)\) core terminals outside
the multiway cut (Lemma~\ref{lem:selection}), avoiding the enumeration
of connecting systems. Using colour coding \cite{gutin2018} and the
reduction of Lemma~\ref{lem:colours}, we optimize over these terminal
sets without enumerating them explicitly. The resulting Steiner Tree
instances have \(O(s)\) terminals and can be solved using the
single-exponential-time, polynomial-space algorithm of Fomin et al.\ \cite{fomin2019}.

\paragraph*{Organization.}
Section~\ref{sec:preliminaries} states the algorithmic ingredients.
Section~\ref{sec:structure} proves the selection identity and its
coloured reduction. Section~\ref{sec:algorithm} gives the algorithm
and its analysis.

\section{Preliminaries}\label{sec:preliminaries}

\subsection{Notation and elementary preprocessing}\label{sec:preprocessing}

For a subgraph \(X\), let \(w(X)\) be the sum of its edge weights.
Deleting a vertex set means taking the induced subgraph on the remaining
vertices. For a weighted graph \(X\) and \(Q\subseteq V(X)\), let
\(\operatorname{OPT}(X,Q)\) be the minimum weight of a connected subgraph
of \(X\) that contains \(Q\). If no such subgraph exists,
\(\operatorname{OPT}(X,Q)=+\infty\). For \(|Q|\leq1\), we use the
convention \(\operatorname{OPT}(X,Q)=0\).

We apply the following preprocessing steps. We contract all zero-weight
edges and update \(K\) and \(S\). We only retain a minimum-weight edge
between each pair of vertices. We return \(0\) if \(|K|\leq1\) and
\(+\infty\) if the terminals lie in different components of \(G\).
Otherwise, we restrict to the component containing all terminals.
This does not influence the possible multiway cuts. Throughout the
analysis, we assume that this preprocessing has been applied; hence
\(G\) is connected, all edge weights are positive, and \(|K|\geq2\).

\subsection{A polynomial-space Steiner Tree algorithm}\label{sec:steiner-solver}

\begin{literaturetheorem}[Fomin et al., {\cite[Theorem~1]{fomin2019}}]
\label{thm:solver}
For a connected graph with \(N\) vertices, \(q\geq 2\) terminals,
and positive integer edge weights at most \(W\), the optimum
Steiner Tree value can be computed deterministically using
\[
O\!\left(7.97^q N^4\log(NW)\right)\text{ time and }
O\!\left(N^3\log(NW)\log q\right)\text{ space}.
\]
\end{literaturetheorem}

Theorem~\ref{thm:solver} gives explicit, low-degree polynomial bounds.
Fomin et al. also give variants with smaller exponential bases and
higher polynomial dependence \cite[Theorems~2 and~3]{fomin2019};
these also suffice for our \(2^{O(s)}\poly(n,b)\) bound.
The theorem numbers for this source refer to the published SIAM version.

For a connected graph \(X\) and required vertices \(Q\), let
\(\STVal(X,Q)\) be \(0\) if \(|Q|\leq1\), and the value returned
by Theorem~\ref{thm:solver} otherwise.

\newpage
\subsection{Deterministic colour coding}\label{sec:colour-coding}

For a positive integer \(a\), write \([a]=\{1,\ldots,a\}\).
The following theorem is obtained by setting \(\alpha=1\) in the
indexed splitter construction of~\cite{gutin2018} and discarding
non-surjective colourings.

\begin{literaturetheorem}[Gutin et al., {\cite[Theorems~4 and~24]{gutin2018}}]
\label{thm:splitter}
For \(2\leq\ell\leq r\), there is a deterministically enumerable
family \(\colourfamily_{r,\ell}\) of surjective functions \(h:[r]\to[\ell]\)
such that, for every \(\ell\)-element set \(A\subseteq[r]\), some
\(h\in\colourfamily_{r,\ell}\) is injective on \(A\).
The family can be enumerated, including initialization, in total
time \(2^{O(\ell)}\poly(r)\) and polynomial space in \(r\).
\end{literaturetheorem}

This construction uses polynomial initialization and query space.
We only store the current colouring. For \(\ell=1\), only the
constant colouring \(h(t)=1\) suffices.

\section{Structural and reduction lemmas}\label{sec:structure}

\begin{definition}[Admissible set]\label{def:admissible}
A set \(U\subseteq S\) is \emph{admissible} if \(U\neq\emptyset\),
\(K\cap S\subseteq U\), and all vertices of \(K\cup U\) belong to
one connected component of \(G-(S\setminus U)\).
\end{definition}

For an admissible set \(U\), we denote this component by \(J_U\) and write
\(\operatorname{OPT}_U=\operatorname{OPT}(J_U,K\cup U)\). Thus the vertices of \(U\)
are required and the vertices of \(S\setminus U\) are forbidden.

\begin{lemma}[Fixing the intersection with the multiway cut]\label{lem:separator}

We have
\begin{equation*}
\operatorname{OPT}(G,K)=\min_{U\text{ admissible}}\operatorname{OPT}_U.
\end{equation*}
\end{lemma}

\begin{proof}
Every connected subgraph of \(J_U\) containing \(K\cup U\) is feasible
for \(K\) in \(G\), so \(\OPT(G,K)\leq\OPT_U\).

Conversely, let \(T\) be an optimal tree for \(K\) in \(G\).
Since \(|K|\geq2\) and each component of \(G-S\) contains at most one
terminal, \(T\) meets \(S\). Set \(U=V(T)\cap S\).
This set contains \(K\cap S\), and \(T\) connects \(K\cup U\) while
avoiding \(S\setminus U\). Thus \(U\) is admissible and \(T\subseteq J_U\),
giving \(\OPT_U\leq w(T)=\OPT(G,K)\).
\end{proof}

For the next two lemmas, fix an admissible set \(U\). We use the notation \(J=J_U\),
\(p=|U|\), \(R=K\setminus S\), and \(r=|R|\). Since \(K\cap S\subseteq U\), we have
\(R\cup U=K\cup U\). Every component of \(J-U\) lies in a component of \(G-S\) and
therefore contains at most one vertex of \(R\). Define
\begin{equation*}
d_t=\operatorname{dist}_J(t,U)\quad(t\in R),\qquad B=\sum_{t\in R}d_t,\qquad D=\max\bigl(\{0\}\cup\{d_t:t\in R\}\bigr).
\end{equation*}
All \(d_t\) are finite. If \(R=\emptyset\), then \(B=D=0\).

The following selection identity builds on the component decomposition
of Jansen and Swennenhuis \cite[Section~2]{jansen2024} and the counting
argument in their Lemma~2.4.

\begin{lemma}[Selection identity]\label{lem:selection}

For \(A\subseteq R\), define
\begin{equation*}
\Phi(A)=B+\operatorname{OPT}(J,U\cup A)-\sum_{t\in A}d_t=\operatorname{OPT}(J,U\cup A)+\sum_{t\in R\setminus A}d_t.
\end{equation*}
Then
\begin{equation*}
\operatorname{OPT}_U=\min_{\substack{A\subseteq R\\ |A|\leq p-1}}\Phi(A).
\end{equation*}
\end{lemma}

\begin{proof}
Fix \(A\subseteq R\) and let \(X\) be an optimal tree for \(U\cup A\)
in \(J\). Add a shortest path from each \(t\in R\setminus A\) to \(U\).
The union is connected and contains \(R\cup U=K\cup U\).
By non-negativity, its weight is at most the sum of the weights of
these parts, so
\begin{equation}\label{eq:upper-selection}
\operatorname{OPT}_U\leq w(X)+\sum_{t\in R\setminus A}d_t=\Phi(A).
\end{equation}

For the reverse inequality, choose an optimal tree \(T\) for \(K\cup U\)
in \(J\) with the fewest edges among all optimal trees. Then
\(w(T)=\OPT_U\), and no proper subgraph of \(T\) connects \(K\cup U\).

For each component \(C\) of \(T-U\), let \(a_C\) count its attachment
edges to \(U\). Since \(T\) is connected and \(U\neq\emptyset\),
we have \(a_C\geq1\). Also, \(T\subseteq J\), so each such component
contains at most one terminal of \(R\). Let \(\mathcal{C}_1\) and
\(\mathcal{C}_{\geq2}\) denote the components with \(a_C=1\) and
\(a_C\geq2\), respectively.

Each \(C\in\mathcal{C}_1\) contains a terminal \(t\in R\): otherwise,
deleting \(C\) and its only attachment edge would contradict the
inclusion-minimality of \(T\). If \(e_C\) is this attachment edge,
then \(C\) together with \(e_C\) contains a path from \(t\) to \(U\),
so \(w(C)+w(e_C)\geq d_t\).

Delete the components in \(\mathcal{C}_1\) and their attachment edges,
and call the remaining subgraph \(H\). It is a tree containing \(U\),
since no component with only one attachment edge lies on a path
between two vertices of \(U\).

Let \(c=|\mathcal{C}_{\geq2}|\), and let \(e_U\) count the edges of \(T\)
with both endpoints in \(U\). Denote by \(H^{\mathrm{contr}}\) the graph
obtained from \(H\) by contracting every component in
\(\mathcal{C}_{\geq2}\) to a single vertex. This tree has \(p+c\)
vertices and \(p+c-1\) edges, giving
\begin{equation*}
2c+e_U\leq\sum_{C\in\mathcal{C}_{\geq2}}a_C+e_U=p+c-1.
\end{equation*}
Since \(e_U\geq0\), we obtain \(c\leq p-1\).
Define \(A=V(H)\cap R\) using the uncontracted tree \(H\).
Each component in \(\mathcal{C}_{\geq2}\) contains at most one terminal
of \(R\), so \(|A|\leq c\leq p-1\).

The deleted components and their attachment edges are pairwise
edge-disjoint and together with \(H\) form \(T\). They contain exactly
the terminals of \(R\setminus A\), one per component, and each contributes
at least the corresponding distance \(d_t\). Hence
\begin{equation*}
w(H)+\sum_{t\in R\setminus A}d_t\leq w(T)=\operatorname{OPT}_U.
\end{equation*}
Since \(H\) connects \(U\cup A\), we have \(\OPT(J,U\cup A)\leq w(H)\)
and thus \(\Phi(A)\leq\OPT_U\). Together with
\eqref{eq:upper-selection}, this proves the identity.
\end{proof}

This lemma shows that \(\operatorname{OPT}_U\) can be obtained by minimizing
the cost of a tree containing \(U\), plus a penalty \(d_t\) for each terminal
\(t\in R\) outside the tree. There exists an optimal tree that needs at most
\(p-1\) core terminals of \(R\). It is possible that, for an arbitrary
choice of \(A\), an optimal tree for \(U\cup A\) contains other terminals
of \(R\setminus A\). This bound allows us to optimize the choice of
these terminals using colour coding.

\begin{figure}[!htbp]
\centering
\includegraphics[width=\linewidth]{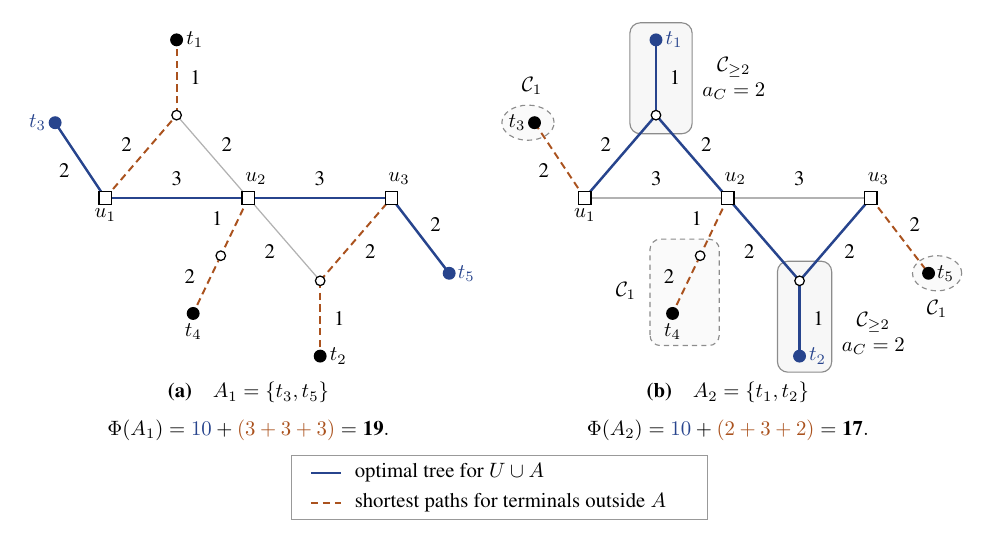}
\par\vspace{-5mm}
\caption{Example with \(K=\{t_1,\ldots,t_5\}\) and \(U=\{u_1,u_2,u_3\}\),
showing two possible choices of \(A\) in \(J_U\).
Blue edges form an optimal tree for \(U\cup A\); dashed orange edges are
shortest paths from the remaining terminals to \(U\).
The value \(\Phi(A)\) is the sum of these weights.
Lemma~\ref{lem:selection} guarantees that some optimal choice satisfies \(|A|\leq p-1\).
In panel (b), the blue and orange edges form an optimal tree \(T\).
The outlined regions are the components of \(T-U\).
Deleting the components in \(\mathcal{C}_1\) and their attachment edges
leaves the blue tree \(H\). The choice \(A_2\) is optimal.}
\label{fig:selection-identity}
\end{figure}

\newpage
\subsection*{Encoding a coloured selection}

We use colour coding to group the non-empty sets \(A\) considered in
Lemma~\ref{lem:selection} into families that can be optimized over together.
Fix \(1\leq\ell\leq\min(p-1,r)\) and assign each terminal in \(R\) one of
\(\ell\) colours, represented by a surjective function \(h:R\to[\ell]\).
Define
\begin{equation*}
\mathcal{F}(h)=\{A\subseteq R:|A|=\ell\text{ and }h(A)=[\ell]\}.
\end{equation*}
The family \(\mathcal{F}(h)\) consists of the sets \(A\) containing
exactly one terminal of each colour.

For each colouring \(h\), we construct a new graph \(J'\) by extending \(J\).
For every colour \(i\in[\ell]\), we add a colour vertex \(z_i\) and connect it
to all terminals \(t\in R\) with \(h(t)=i\).
The edge \(z_it\) receives weight \(M-d_t\), where
\begin{equation*}
M=\Lambda+(\ell+1)D+1,
 \qquad \Lambda=\sum_{e\in E(J)}w(e).
\end{equation*}
Let \(Q=U\cup\{z_i:i\in[\ell]\}\) be the required vertex set in \(J'\).
The discount \(d_t\) removes the contribution of a selected terminal
from the baseline \(B\).

\begin{lemma}[Coloured selection reduction]\label{lem:colours}

Writing \(F=\operatorname{OPT}(J',Q)\), we have
\begin{equation*}
F-\ell M
 =\min_{A\in\mathcal{F}(h)}
   \left[\operatorname{OPT}(J,U\cup A)-\sum_{t\in A}d_t\right].
\end{equation*}
\end{lemma}

\begin{proof}
Since \(h(R)=[\ell]\), every new colour vertex has a neighbour in \(J\),
so \(J'\) is connected.

We first show that the large base cost \(M\) forces all new colour vertices
to have degree one in every optimal tree for \(Q\) in \(J'\).

Take a spanning tree of \(J\) and attach each of the \(\ell\) colour vertices
to this tree using one of its incident edges. This gives a tree in \(J'\)
containing \(Q\) with cost at most \(\Lambda+\ell M\). Thus,
\begin{equation*}
F\leq\Lambda+\ell M.
\end{equation*}
Suppose that an optimal Steiner tree \(T'\) for \(Q\) in \(J'\) uses at least
\(\ell+1\) new edges. Each new edge has weight at least \(M-D\), so
\begin{equation*}
w(T')\geq(\ell+1)(M-D)=\ell M+\Lambda+1>\ell M+\Lambda\geq F.
\end{equation*}
This contradicts \(w(T')=F\). Every colour vertex needs an incident edge,
so \(T'\) uses exactly \(\ell\) new edges and every \(z_i\) has degree one in \(T'\).

Let \(t_i\) be the neighbour of \(z_i\) in \(T'\). Since \(h(t_i)=i\) for every
\(i\in[\ell]\), these vertices have different colours and are therefore distinct. Hence
\(A=\{t_i:i\in[\ell]\}\) belongs to \(\mathcal{F}(h)\). Removing the colour leaves from
\(T'\) yields a tree \(X\) in \(J\) containing \(U\cup A\). Therefore,
\begin{equation*}
\begin{aligned}
 F&=w(X)+\ell M-\sum_{t\in A}d_t\\
  &\geq\operatorname{OPT}(J,U\cup A)+\ell M-\sum_{t\in A}d_t.
\end{aligned}
\end{equation*}
Thus \(F-\ell M\) is at least the minimum on the right-hand side
of the claimed identity.

Conversely, take any \(A\in\mathcal{F}(h)\) and an optimal tree for \(U\cup A\) in
\(J\). For each \(i\in[\ell]\), let \(t_i\) be the unique terminal in \(A\) with colour
\(i\). Attach \(z_i\) to the tree by adding only the edge \(z_it_i\) from \(J'\).
This gives a tree in \(J'\) containing \(Q\), of weight
\begin{equation*}
\operatorname{OPT}(J,U\cup A)+\ell M-\sum_{t\in A}d_t.
\end{equation*}
Taking the minimum over \(A\in\mathcal F(h)\) gives the reverse inequality.
\end{proof}

\section{The algorithm and its analysis}\label{sec:algorithm}

\subsection{Computing the optimum value}\label{sec:value}

This section describes the algorithm \(\ST(G,K,S)\).
Every call to \(\ST\) first applies the preprocessing from
Section~\ref{sec:preprocessing}; Algorithm~\ref{alg:value} describes the
remaining case. It uses \(\STVal\) as defined in Section~\ref{sec:steiner-solver}.
Use the families \(\colourfamily_{r,\ell}\) from
Section~\ref{sec:colour-coding}, identifying \(R\) with \([r]\).

\begin{algorithmblock}
\refstepcounter{algorithm}\label{alg:value}
\textbf{Algorithm \thealgorithm.} \(\ST(G,K,S)\)
\par\smallskip
\textbf{Input:} A weighted graph \(G\), terminals \(K\), and a multiway cut \(S\).\\
\textbf{Output:} \(\OPT(G,K)\), or \(+\infty\) if infeasible.
\begin{tabbing}
\hspace{1.3em}\=\hspace{1.3em}\=\hspace{1.3em}\=\hspace{1.3em}\=\kill
\(\mathit{best}\gets+\infty\)\\
\textbf{for each} admissible set \(U\subseteq S\), enumerated one at a time, \textbf{do}\\
\>Construct \(J=J_U\); set \(p=|U|\), \(R=K\setminus S\), and \(r=|R|\)\\
\>Compute \(d_t\) for all \(t\in R\), and \(B,D,\Lambda\)\\
\>\(\mathit{best}\gets\min\{\mathit{best},B+\STVal(J,U)\}\) \quad (the \(\ell=0\) case)\\
\>\textbf{for} \(\ell=1,\ldots,\min(p-1,r)\) \textbf{do}\\
\>\>\(M\gets\Lambda+(\ell+1)D+1\)\\
\>\>\textbf{for each} \(h\in\colourfamily_{r,\ell}\) \textbf{do}\\
\>\>\>Construct \((J',Q)\) as in Lemma~\ref{lem:colours}\\
\>\>\>\(F\gets\STVal(J',Q)\)\\
\>\>\>\(\mathit{best}\gets\min\{\mathit{best},B+F-\ell M\}\)\\
\>\>\textbf{end for}\\
\>\textbf{end for}\\
\textbf{end for}\\
\textbf{return} \(\mathit{best}\)
\end{tabbing}
\end{algorithmblock}

\subsection{Correctness}\label{sec:correctness}

\begin{proposition}\label{prop:correct}
The procedure \(\ST(G,K,S)\) returns \(\OPT(G,K)\) if a Steiner
tree for \(K\) in \(G\) exists, and \(+\infty\) otherwise.
\end{proposition}

\begin{proof}
The preprocessing from Section~\ref{sec:preprocessing} handles the trivial
and infeasible cases and preserves the optimum otherwise.
Consider a non-trivial instance. Fix an admissible \(U\) and let \(b_U\)
be the minimum candidate generated for \(U\).

The \(\ell=0\) candidate is \(\Phi(\emptyset)\geq\OPT_U\).
For every pair \((\ell,h)\) with \(\ell>0\),
Lemmas~\ref{lem:colours} and~\ref{lem:selection} give
\begin{equation*}
B+F-\ell M=\min_{A\in\mathcal{F}(h)}\Phi(A)\geq\operatorname{OPT}_U.
\end{equation*}
Hence \(b_U\geq\operatorname{OPT}_U\).

By Lemma~\ref{lem:selection}, some \(A^*\subseteq R\) satisfies
\(|A^*|\leq p-1\) and \(\Phi(A^*)=\OPT_U\).
If \(A^*=\emptyset\), equality is attained by the \(\ell=0\)
candidate. Otherwise, take \(\ell=|A^*|\).
Theorem~\ref{thm:splitter} (or the constant colouring for \(\ell=1\))
provides at least one colouring \(h^*\) that is injective on \(A^*\).
Since \(|A^*|=\ell\), we have \(h^*(A^*)=[\ell]\), and hence
\(A^*\in\mathcal F(h^*)\). The candidate for \(h^*\) therefore satisfies
\begin{equation*}
B+F-\ell M
 =\min_{A\in\mathcal{F}(h^*)}\Phi(A)
 \leq\Phi(A^*)=\operatorname{OPT}_U.
\end{equation*}
Together with \(b_U\geq\operatorname{OPT}_U\), this proves
\(b_U=\OPT_U\).

It follows from Lemma~\ref{lem:separator} that the returned value is
\begin{equation*}
\min_{U\text{ admissible}}b_U
 =\min_{U\text{ admissible}}\operatorname{OPT}_U
 =\operatorname{OPT}(G,K).
\end{equation*}
\end{proof}

\subsection{Running time}\label{sec:time}

The preprocessing takes polynomial time. Consider an instance with \(s\geq1\).
For each of the \(2^s\) choices of \(U\subseteq S\), checking admissibility
and, if \(U\) is admissible, constructing \(J_U\) and computing distances
to \(U\) in \(J_U\) take polynomial time. There is one solver call for
\(\ell=0\).

For fixed \(\ell>0\), enumerating \(\colourfamily_{r,\ell}\), including
initialization, takes \(2^{O(\ell)}\poly(n)\) time
(Section~\ref{sec:colour-coding}). Each colouring requires one solver call.

To bound the cost of each solver call, define
\begin{equation*}
W=\max\bigl(\{1\}\cup\{w(e):e\in E\}\bigr),
 \qquad \beta=b+\lceil\log(n+m+2)\rceil.
\end{equation*}
For an admissible \(U\) and \(\ell>0\), constructing \(J'\) from \(J\)
adds \(\ell\) colour vertices and one new edge for each \(t\in R\).
Since \(p\leq s\),
\begin{equation*}
|V(J')|\leq2n,\qquad |E(J')|\leq m+n,\qquad |Q|=p+\ell\leq2s-1.
\end{equation*}

Since \(D\leq(n-1)W\) and \(\ell+1\leq s\), every new edge has
weight at most \(M\), where
\begin{equation*}
M=\Lambda+(\ell+1)D+1\leq mW+s(n-1)W+1.
\end{equation*}
This bound has \(O(\beta)\) bits. Constructing \(J'\) therefore takes
polynomial time in \(n\) and \(b\).
Theorem~\ref{thm:solver} bounds every solver call \(\STVal(J',Q)\) by
\begin{equation*}
 7.97^{p+\ell}\poly(n,b)
 \leq 7.97^{2s-1}\poly(n,b).
\end{equation*}
Combining these bounds, the overall running time is bounded by
\begin{equation*}
\begin{aligned}
 &2^s\left[
  7.97^s+
  \sum_{\ell=1}^{s-1}2^{O(\ell)}\,7.97^{2s-1}
 \right]\poly(n,b)\\
 &\hspace{2em}=2^{O(s)}\poly(n,b).
\end{aligned}
\end{equation*}
Polynomial factors in \(s\) are absorbed into \(\poly(n,b)\),
since \(s\leq n\).

\subsection{Polynomial space}\label{sec:space}

Since \(B\leq r(n-1)W\) and \(F\leq(|V(J')|-1)M\), all stored
distances and finite costs, including \(B,F,\ell M\), and \(B+F-\ell M\),
have \(O(\beta)\) bits. Table~\ref{tab:space} lists the data stored
during one iteration.

\begin{spacetable}
\caption{Storage used by the current iteration and one active solver call.
The polynomial \(P_{\mathrm{split}}(n)\) bounds the space used by the
splitter construction in Section~\ref{sec:colour-coding}.}
\label{tab:space}
\small
\renewcommand{\arraystretch}{1.25}
\begin{tabularx}{\linewidth}{@{}X X@{}}
\toprule
Stored data & Bit-space bound\\
\midrule
Graphs \(G\), \(J\), and the current \(J'\)
 & \(O((n+m)\beta)\)\\
Set \(U\) and loop counters
 & \(O(n)\)\\
One colouring \(h:R\to[\ell]\)
 & \(O(r\log(\ell+1))\)\\
Distances, scalar costs, and running minimum
 & \(O(n\beta)\)\\
Splitter
 & \(P_{\mathrm{split}}(n)\)\\
One active call to Theorem~\ref{thm:solver}
 & \(O(n^3\beta\log(n+2))\)\\
\bottomrule
\end{tabularx}
\end{spacetable}

The current set \(U\) uses \(s\) bits.
The indexed generator of Theorem~\ref{thm:splitter} uses polynomial
initialization and query space, with
\(O(\ell+\log(n+2))\) bits for its current index.
Only one colouring is stored during its solver call.
Checking that all colours occur uses \(O(\ell)\) additional bits.

Summing the storage used simultaneously gives
\begin{equation*}
\begin{aligned}
 \Space(\ST)
 &\leq P_{\mathrm{split}}(n)
    +O\bigl((n+m)\beta+n^3\beta\log(n+2)\bigr)\\
 &=\poly(n,b).
\end{aligned}
\end{equation*}
The degrees of these polynomial bounds are independent of \(s\).

\subsection{Recovering an optimal tree}\label{sec:reconstruction}

Let \(G,K,S\) denote the original feasible instance with \(|K|\geq2\),
before preprocessing. Set \(v=\ST(G,K,S)\) and \(G_{\mathrm{cur}}=G\).
Process each original edge \(e\) once, deleting it from \(G_{\mathrm{cur}}\)
if \(\ST(G_{\mathrm{cur}}-e,K,S)=v\).
An infeasible call returns \(+\infty\), so the edge is retained.

The deletions preserve the optimum \(v\).
Any edge outside an optimal tree of the final graph would have been
deleted when tested, since this tree was already feasible then.
Thus, after discarding isolated non-terminals, the final graph is a
tree \(T\) with \(w(T)=v\).

Edge deletion preserves \(S\) as a multiway cut.
Reconstruction makes at most \(m\) additional calls to \(\ST\)
and reuses the same workspace. Thus the time bound of
\(2^{O(s)}\poly(n,b)\) and the space bound of \(\poly(n,b)\) remain,
proving Theorem~\ref{thm:main}.

\newpage
\section{Conclusion}\label{sec:conclusion}

This paper answers the question of Jansen and Swennenhuis by giving an
algorithm for Steiner Tree that runs in \(2^{O(s)}\poly(n,b)\) time and
polynomial space, where \(s\) is the size of a multiway cut for the
terminals. The selection identity reduces the optimization for each
admissible \(U\subseteq S\) to choosing at most \(|U|-1\) core terminals
outside the multiway cut. Using colour coding and the weighted reduction,
we optimize over these choices without enumerating all such sets explicitly.

A further question is whether a \(2^{o(s)}\poly(n,b)\)-time algorithm
exists for Steiner Tree. Such an algorithm would contradict the
Exponential Time Hypothesis: taking \(S=K\) would yield a
\(2^{o(k)}\poly(n,b)\)-time algorithm, contrary to the known
ETH-based lower bound \cite[Theorem~I.2]{marx2018}.
The recent proposed refutation of ETH \cite{openai2026eth} gives
motivation to investigate this question, although the exponential factors
in our algorithm suggest that a different approach may be needed.

\section*{Use of generative AI}

GPT-Astra (OpenAI) was asked to work on the open question posed by
Jansen and Swennenhuis \cite{jansen2024}. The model contributed substantially
to the development of the algorithm and the proofs through an iterative
dialogue with the first author. It was also used extensively to improve
the exposition by rewriting and clarifying the mathematical arguments.
The authors take full responsibility for the content of the paper.

\bibliography{references}
\end{document}